\documentclass[1p,times,procedia]{elsarticle}
\usepackage{nupha_ecrc}
\usepackage{float}

\volume{00}

\firstpage{1}

\journalname{Nuclear Physics A}

\runauth{C.~Plumberg and U.~Heinz}

\jid{nupha}

\jnltitlelogo{Nuclear Physics A}

\usepackage{amssymb}

\usepackage{lineno}

\biboptions{comma,sort&compress}

\usepackage[figuresright]{rotating}

	\newcommand{\avg}[1]{\left< #1 \right>}
	\newcommand{\shortavg}[1]{\langle #1 \rangle}
	\def\l{\left}
	\def\r{\right}
	\def\ev{\mathrm{ev}}
	
	\def\mcO{\mathcal{O}}
	\def\mcP{\mathcal{P}}
	
	\newcommand{\eqref}[1]{(\ref{#1})}

\begin{document}

\begin{frontmatter}



\dochead{}

\title{Observable consequences of event-by-event fluctuations of HBT radii}


\author{Christopher Plumberg and Ulrich Heinz}

\address{Physics Department, The Ohio State University, 191 W. Woodruff Ave, Columbus, OH 43210}

\begin{abstract}
We explore the effects of event-by-event fluctuations of Hanbury Brown--Twiss (HBT) radii and show how they can be observed experimentally. The relation of measured HBT radii extracted from ensemble-averaged correlation functions to the mean of their event-by-event probability distribution is clarified. We propose a method to experimentally determine the mean and variance of this distribution and test it on an ensemble of fluctuating events generated with the viscous hydrodynamic code {\sc VISH2+1}. Using the same code, the sensitivity of the mean and variance of the HBT radii to the specific QGP shear viscosity $\eta/s$ is studied. We report sensitivity of the mean pion HBT radii and their variances to the temperature dependence of $\eta/s$ near the quark-hadron transition at a level similar (10-20\%) to that which was previously observed for elliptic and quadrangular flow of charged hadrons \cite{Molnar:2014zha}.
\end{abstract}


\end{frontmatter}


\section{Introduction}
\label{Introduction}

For the past several years, event-by-event fluctuations in the initial conditions of heavy-ion collisions have played a critical role in our understanding of how these collisions evolve. Their manifestations have been studied extensively by means of a wide variety of momentum-space observables, such as the mean transverse momenta $\avg{p_T}$ and the anisotropic flow coefficients $v_n$, which all fluctuate from event to event. The fluctuations of these observables are characterized by probability distributions over a set of collision events with tightly constrained collision geometries; different moments of these distributions are accessible with different observables. 

We here extend the study of these fluctuations by exploring their effect on HBT interferometry. In contrast to flow measurements, which reflect the momentum-space structure of the final state of a heavy-ion collision, measurements of HBT radii provide insight into its space-time structure. By finding a way to characterize the event-by-event distribution of fluctuating HBT radius parameters with a set of measurements that determine a few of its moments, we hope to obtain insights into the fluctuations of the size and shape of the collision fireballs at freeze-out. Theoretical modeling of these HBT fluctuations can establish to what extent measuring their moments can provide additional information about the properties of the medium whose evolution connects the observed final state with the fluctuating initial state -- information that, being based on a spatio-temporal measurement of the final state, may be complementary to that extracted from flow measurements.   

\section{Formalism}
\label{Formalism}

For an ensemble of events, the normalized and ensemble-averaged two-particle correlation function, for a pair of momenta $\vec{p}_1$ and $\vec{p}_2$, is defined experimentally by
\begin{equation}
C_{\mathrm{avg}}(\vec{p}_1, \vec{p}_2) \equiv \frac{\l< E_{p_1} E_{p_2} \frac{d^6 N}{d^3 p_1 d^3 p_2} \r>_{\ev}}{ \l<E_{p_1} \frac{d^3 N}{d^3 p_1}\r>_{\ev} \l< E_{p_2} \frac{d^3 N}{d^3 p_2} \r>_{\ev} } .\label{corrfuncENSAVG0defn}
\end{equation}  
Using the Gaussian source approximation to write the HBT radii in terms of space-time variances of the source \cite{Heinz:2004qz,Heinz:1999rw,Wiedemann:1999qn}, the radii associated with the correlation function \eqref{corrfuncENSAVG0defn} can be shown \cite{Plumberg:2015mxa} to be an event-multiplicity-weighted average of the HBT radii of the individual events in the ensemble:
\begin{equation}
R^2_{\avg{ij}}(\vec{K}) \equiv \avg{w R^2_{ij}}_{N_\ev} 
	\equiv \frac{\sum^{N_{\ev}}_{k=1} N^2_k(\vec{K}) \l( R^2_{ij}(\vec{K})\r)_k}{\sum^{N_{\ev}}_{k=1} N^2_k(\vec{K})},
			\ \  {\mbox{where\ \ }}N_k(\vec{K}) \equiv \left(E_K \frac{d^3 N}{d^3 K}\right)_k. \label{R2ij_from_Cavg}
\end{equation}  
$N_k(\vec{K})$ is the multiplicity of particles with momentum $\vec{K}$ in the $k$th event, and the weights in the average represent the number of pairs with pair momentum $\vec{K}$ contributed by each event to the ensemble-averaged correlation function.  We refer to these ensemble-averaged radii as their \textit{physical ensemble average} (PEA); this PEA is measured in the experiments. The PEA differs in general from the simple arithmetic average of the radius parameters in the measured ensemble,
\begin{equation}
\label{eq3}
   \avg{R^2_{ij}(\vec{K})}_{N_\ev} \equiv \frac{1}{N_{\ev}} \sum^{N_{\ev}}_{k=1} \l ( R^2_{ij}(\vec{K})\r)_k,
   \ \  {\mbox{where\ \ }}\avg{\mcO}_{N_\ev} \equiv \frac{1}{N_{\ev}} \sum^{N_{\ev}}_{k=1} \mcO_k\,.
\end{equation}  
We refer to Eq.~(\ref{eq3}) as the \textit{direct ensemble average} (DEA).  Denoting the distribution of event-by-event radii for an ensemble of events formally by $\mcP_{N_{\ev}}(R^2_{ij})$, the DEA radii represent its \textit{true} mean whereas the PEA radii represent a weighted mean. No technique exists currently to correct the measured PEA radii to obtain a good estimate for the DEA radii.

To further characterize $\mcP_{N_{\ev}}(R^2_{ij})$ (which cannot be measured directly because limited pair multiplicities make a full HBT analysis of individual events impossible), we can try to determine also a few of its higher statistical moments, e.g., its variance $\sigma^2_{ij,N_{\ev}} \equiv {\mbox{Var}}\l[ \mcP_{N_{\ev}}(R^2_{ij}) \r]$. In \cite{Plumberg:2015mxa} we have proposed the following set of techniques for experimentally constructing estimates for both the DEA mean $\avg{R^2_{ij}}_{N_{\ev}}$ and the variance $\sigma^2_{ij,N_{\ev}}$ of the ensemble distribution of radii $\mcP_{N_{\ev}}(R^2_{ij})$:
\noindent\begin{enumerate}
\item
To estimate the DEA radii for an ensemble containing $N_\ev$ events, we sort the events by increasing multiplicity $N(\vec{K})$, split them into $\tilde{n}_b$ bins of $\tilde{n} \equiv {N_{ev}} / \tilde{n}_b$ events each, and compute the PEA radii $\avg{w \mcO}^{(\ell)}_{\tilde{n}}$ for the sub-ensembles corresponding to each bin $\ell=1,\dots,\tilde{n}_b$.  If $n_b$ is sufficiently large (but still $n_b{\,\ll\,}N_{\ev}$), the weights may be treated as approximately constant within each bin, implying that $w^{(\tilde{n})}_{k} \approx \frac{\tilde{n}_b}{N_{ev}}$. With this approximation one shows that \cite{Plumberg:2015mxa}
\begin{equation}
\avg{\mcO}_{N_{ev},{\mathrm{est}}} \equiv \frac{1}{\tilde{n}_b} \sum^{\tilde{n}_b}_{\ell=1} \avg{w \mcO}^{(\ell)}_{n}
	\approx \avg{\mcO}_{N_\ev}.
\label{bin_average_approximate_mean}
\end{equation}
\item
To estimate the variance, we begin again with an ensemble containing $N_\ev$ events, and we consider an iterative procedure consisting of the following two steps.  For the $k$th iteration, we first split the $N_\ev$ events into $n_b$ equally-sized sub-ensembles, and estimate the DEA radii $\avg{R^2_{ij}}_{k,\ell}$ for each of these sub-ensembles ($\ell = 1,\dots,n_b$).  We then compute the variance of these $n_b$ estimates for the DEA radii, which in turn yields an estimate for the variance of the distribution of DEA radii of sub-ensembles of size $n \equiv N_\ev/n_b$.  Iterating these steps $M$ times thus generates $M$ variance estimates which we then average and rescale by a factor determined by the central limit theorem, finally obtaining the following estimate for $\sigma^2_{ij,N_{\ev}}$:
\begin{equation}
	\sigma^2_{ij, N_\ev, {\mathrm{est}}}
		\equiv \frac{N_{\ev}/n_b}{M (n_b{-}1) }
				\sum_{k=1}^M \sum_{\ell=1}^{n_b}
						\l( \avg{R^2_{ij}}_{k,\ell}^2 - \avg{R^2_{ij}}_{N_{\ev}}^2 \r) .\label{variance_estimator}
\end{equation}
\end{enumerate}
\section{Results}
\label{Results}
\begin{figure}[H]
\hspace*{-2.5mm}
\begin{tabular}{@{}c@{}}
	\begin{tabular}{@{}c@{}}
		\includegraphics[width=0.425\linewidth]{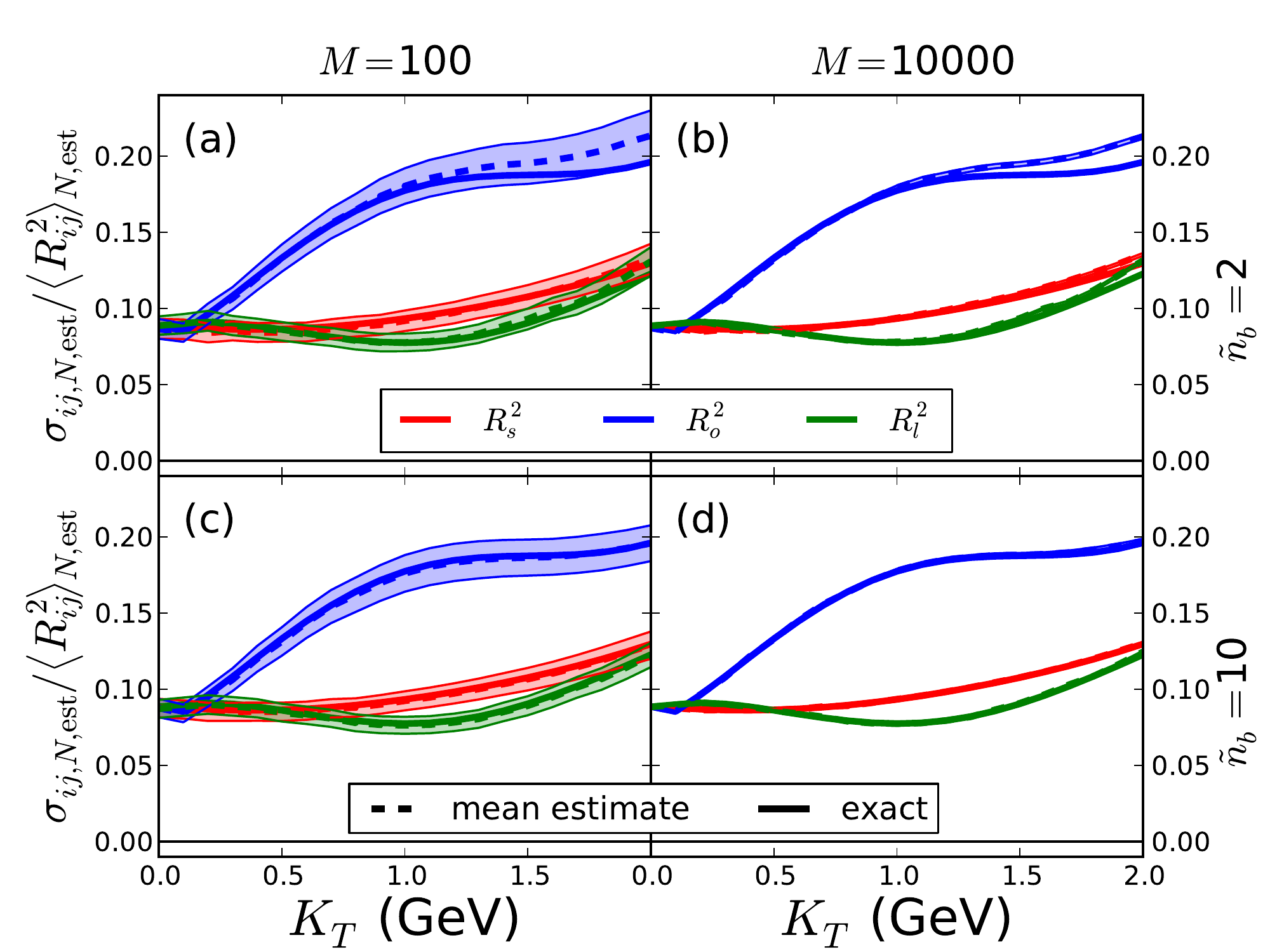}
		\\
		\includegraphics[width=0.425\linewidth]{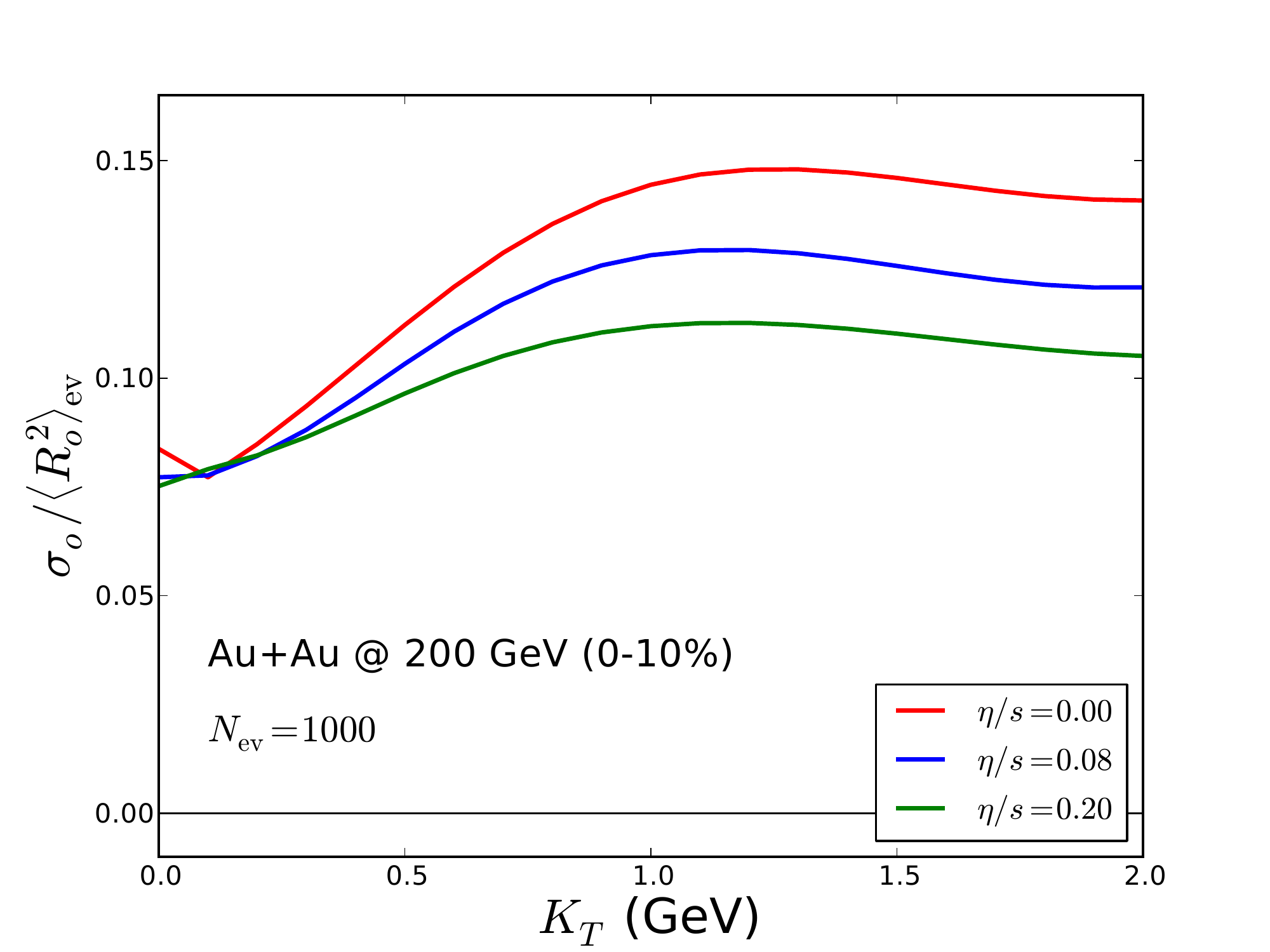}%
	\end{tabular}
	\hspace*{-0.5mm}
	\begin{tabular}{@{}c@{}}
		\includegraphics[width=0.58\linewidth]{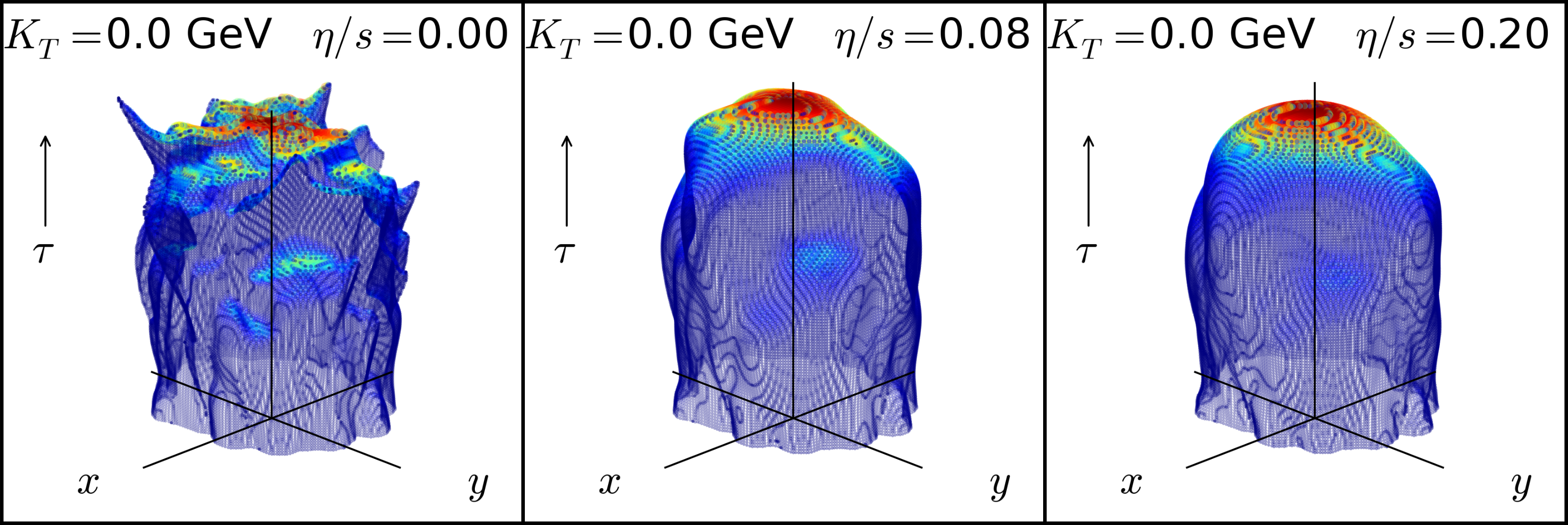}
		\\
		\includegraphics[width=0.58\linewidth]{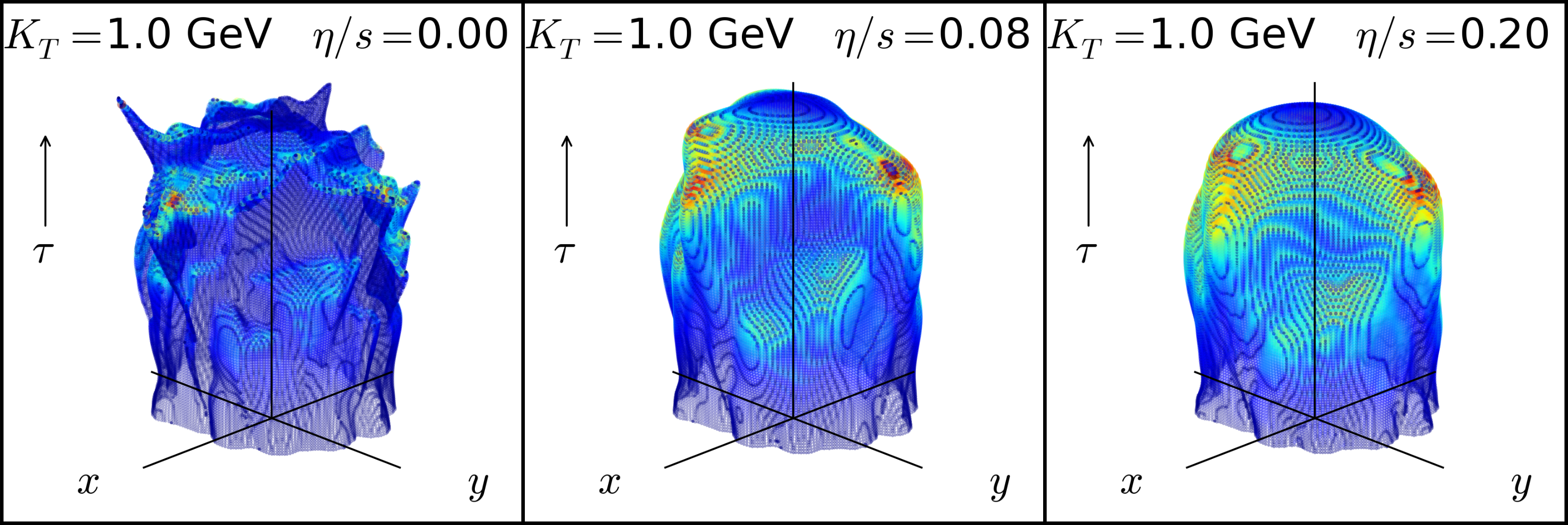}%
		\\
		\includegraphics[width=0.58\linewidth]{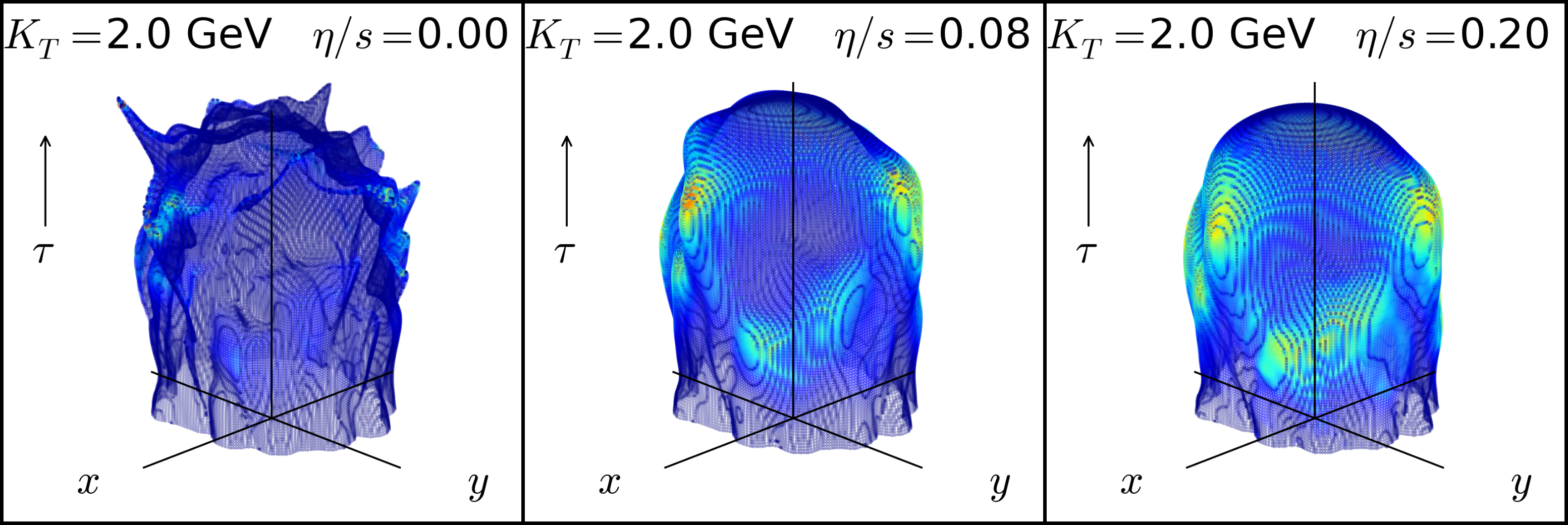}%
	\end{tabular}
\end{tabular}
\caption{{\bf Top left:} Comparison of estimated relative widths 
              $\sigma_{ij,N_{\ev}}/\shortavg{R^2_{ij}}_{N_{\ev}}$ with their exact values as 
              functions of $K_T$, for 5000 0-10\% Au+Au events at 200$A$ GeV, for 
              $M=100$, $10000$ (left vs. right) and $\tilde{n}_b=2$, 10 (top vs. bottom), 
              using $n_b=2$.
              {\bf Bottom left:} Exact relative widths $\sigma_{o, N_{\ev}}/\shortavg{R^2_{o}}_{N_{\ev}}$
              as function of $K_T$ for 1000 0-10\% Au+Au events at 200$A$ GeV, 
              using $(\eta/s)_{QGP} = 0.00$, 0.08, and 0.20 as indicated.		
              {\bf Right:} Effective emission regions on the fluctuating freeze-out surface for a single 
              event with fixed bumpy initial conditions, for different values of the pair momentum 
              $K_T$ (rows) and for hydrodynamic evolution with different $(\eta/s)_{QGP}$ values 
              (columns).
\label{Fig1}}
\end{figure}

We now use these techniques to estimate the relative widths $\sigma_{ij}/\shortavg{R^2_{ij}}_{N_{\ev}}$ of an ensemble of $N_{\ev}{\,=\,}5000$ 200$A$\,GeV Au+Au events at 0-10\% centrality.  Further details of our analysis are provided in \cite{Plumberg:2015mxa,Plumberg:2015eia}.  Fixing $n_b = 2$ in Eq.~\eqref{variance_estimator}, we obtain the estimates for the relative widths shown in the top left part of Fig.~\ref{Fig1}.  We see that increasing $\tilde{n}_b$ decreases the bias of our estimator, while increasing $M$ decreases the variability and improves the statistical precision.

To illustrate possible useful applications of our proposed techniques we demonstrate the sensitivity of the relative widths of $R^2_o$ $\l(\sigma_o/\shortavg{R^2_o}_{N_{\ev}}\r)$ to the value of the specific shear viscosity $\eta/s$ in the quark-gluon plasma (QGP) phase of heavy-ion collision evolution. These results are shown in the lower lefthand panel of Fig.~\ref{Fig1}.  The observed increase of $\sigma_o/\shortavg{R^2_o}_{N_{\ev}}$ at large $K_T$ can be shown \cite{Plumberg:2015eia} to result primarily from event-by-event fluctuations of the emission duration, not the emission geometry, while the decrease with increasing $\eta/s$ can be understood as a suppression of event-by-event fluctuations due to the increased dissipation in the hydrodynamic evolution.

This intuition is confirmed in the right part of Fig.~\ref{Fig1} where we show, for three values of the pair momentum $\vec{K}$ and three different choices of the specific shear viscosity $\eta/s$ in the hydrodynamic evolution, the regions of greatest emissivity along the freeze-out surface for a typical fluctuating event in our ensemble. Red regions represent strongest particle emission, and blue regions the weakest.  For small $K_T$ (top row), the effective emission regions are clearly centered at small radii and late times, while for larger $K_T$ values (middle and bottom rows) the effective emission regions shift toward larger radii and earlier times. As the freeze-out surface fluctuates from event to event, due to the generic shape of the freeze-out surface the emission regions at large $K_T$ are constrained to fluctuate primarily in the temporal (vertical) direction and not much in the radial (transverse) direction. This implies that, at large $K_T$, fluctuation signatures will tend to be dominated by emission duration over emission geometry, as stated above.  Moreover, a comparison of the ideal results in the lefthand column ($\eta/s = 0$) with the viscous results in the middle and righthand columns ($\eta/s = 0.08$ and $0.2$) reveals a clear suppression of the bumpiness of the freeze-out surface as $\eta/s$ is increased, leading to the reduced variance of the HBT radii seen in the lower left panel of Fig.~\ref{Fig1}. We conclude that the new observables proposed here can indeed yield valuable additional insights into heavy-ion collisions and their evolution.

\noindent \textbf{Acknowledgements:} We thank Jai Salzwedel and Michael Lisa for valuable discussions. This work was supported by the Department of Energy, Office of Science, Office of Nuclear Physics under Awards No. \rm{DE-SC0004286} and (within the framework of the JET Collaboration) \rm{DE-SC0004104}. Computing resources provided by the Ohio Supercomputer Center \cite{OhioSupercomputerCenter1987} are gratefully acknowledged.
%
%
%




\bibliographystyle{elsarticle-num}
\bibliography{Plumbergbib}







\end{document}